\documentclass[twocolumn]{aastex63}

\received{June 3, 2021}
\accepted{\today}

\usepackage[utf8]{inputenc}
\usepackage[T1]{fontenc}

\usepackage{graphicx}	
\usepackage{amsmath}	
\usepackage{amssymb}	
\usepackage{multirow}
\usepackage{booktabs}
\usepackage{textcomp}
\usepackage{hyperref}
\hypersetup{colorlinks, linkcolor={red}, citecolor={blue}, urlcolor={magenta}}
\usepackage{morefloats}
\usepackage[separate-uncertainty,multi-part-units = single]{siunitx}
\usepackage{tabularx}
\usepackage{natbib}

\newcommand{\ind}[2]{\ensuremath{#1_\mathrm{#2}}}

\DeclareSIUnit\solarmass{\ensuremath{M_\odot}}
\DeclareSIUnit\solarlum{\ensuremath{L_\odot}}
\DeclareSIUnit\angstrom{\ensuremath{\mathrm{\mbox{\AA}}}}
\DeclareSIUnit\parsec{\ensuremath{\mathrm{pc}}}
\DeclareSIUnit\erg{\ensuremath{\mathrm{erg}}}
\DeclareSIUnit\year{\ensuremath{\mathrm{yr}}}
\defcitealias{2015ApJ...800L..35A}{AP15} 

\shorttitle{}
\shortauthors{Savić et al.}

\begin{document}

\title{The first supermassive black hole mass measurement in active galactic nuclei using the polarization of broad emission line Mg\,II}

\correspondingauthor{Đ.\,Savić}
\email{djsavic@aob.rs}

\author[0000-0003-0880-8963]{Đorđe V.\,Savić}
\affiliation{Institut d’Astrophysique et de Géophysique, Université de Liège \\ Allée du 6 Août 19c, 4000 Liège, Belgium}
\affiliation{Astronomska opservatorija, Beograd \\
Volgina 7, 11060 Belgrade, Serbia}

\author[0000-0003-2398-7664]{Luka Č.\,Popović}
\affiliation{Astronomska opservatorija, Beograd \\
Volgina 7, 11060 Belgrade, Serbia}
\affiliation{Department of Astronomy, Faculty of Mathematics, University of Belgrade\\
             Studentski trg 16, 11000 Belgrade, Serbia}

\author[0000-0003-2914-2507]{Elena Shablovinskaya}
\affiliation{Astrophysical Observatory of the Russian Academy of Sciences \\
             Nizhnij Arkhyz, Karachaevo-Cherkesia 369167, Russia
            }




\begin{abstract}

Spectropolarimetric efforts in the last few years have provided an efficient method that is based on the profiles of the polarization plane position angle of broad emission lines in active galactic nuclei (AGNs). Here we present black hole measurements of SBS\,1419+538 using spectropolarimetric observations in the Mg\,II spectral band. The observations are performed by 6m telescope of SAO RAS using SCORPIO-2. We found a good agreement for the estimated supermassive black hole (SMBH) mass for this object using spectropolarimetry when compared with the mass obtained using other methods.
\end{abstract}



\section{Introduction} \label{s:intro}
According to the standard paradigm, AGNs are powered by an accretion of gas onto the SMBH which resides in the center \citep{1964ApJ...140..796S,1964SPhD....9..246Z,1969Natur.223..690L}. Due to finite, but low viscosity of the gas, the gas temperature increases and the angular momentum is being transfered outwards, providing a slow, but steady inflow \citep{1973A&A....24..337S}. The gravitational binding energy is converted into enormous amount of radiation which ranks AGNs as the most luminous steady sources observed \citep{2017FrASS...4...35P}. SMBHs actively shape the environment in its vicinity, but also on \SI{}{\kilo\parsec} scales, through a process known as AGN feedback \citep{2012ARA&A..50..455F}, which plays an important role in the host galaxy evolution \citep{2013ARA&A..51..511K,2014ARA&A..52..589H}. Therefore, reliable SMBH mass estimation is an important problem in modern astrophysics.

Many methods with different approach have been developed in the past few decades and extensively discussed in the literature \citep[e.g.][and references therein]{2014SSRv..183..253P,2020OAst...29....1P}. Broad emission lines in AGNs have been widely used for measuring SMBHs mass, most often in a long term reverberation mapping campaigns \citep[][etc.]{1982ApJ...255..419B,1993PASP..105..247P,2000ApJ...533..631K,2013ApJ...767..149B,2016ApJ...820...27D,2018ApJ...856....6D,2019MNRAS.485.4790S}.

When the polarized emission is taken into account, the broad line spectropolarimetry allows us to measure the SMBH mass with a single-epoch observations \citep[][hereafter \citetalias{2015ApJ...800L..35A} method,]{2015ApJ...800L..35A}. This method assumes that equatorial scattering of the inner side of the dusty torus is the dominant polarization mechanism \citep{2005MNRAS.359..846S,2018A&A...614A.120S,2020MNRAS.497.3047S,2020MNRAS.491....1L} and is in a good agreement with other methods \citep{2019MNRAS.482.4985A}. In order to use the \citetalias{2015ApJ...800L..35A} method, it is required that the distance ($R_{sc}$) between the SMBH and the scattering region is known. 
In the case of  the Keplerian-like motion in combination with the equatorial scattering of the BLR
light, the relation between velocities  and polarization angle 
($\tan\varphi$) across the broad line is:

\begin{equation}
\log\left(\frac{V_i}{c}\right)=a-b\cdot \log\left(\tan(\Delta\varphi_i)\right)
\end{equation}
where $c$ is the speed of light, and constant $a$ depends on the BH mass \ind{\mathcal{M}}{bh} as
\begin{equation}
a=0.5\log\left({\frac{G\ind{\mathcal{M}}{bh} \cos^2(\theta)}{c^2\ind{R}{sc}}}\right),
\end{equation}
where $G$ is the gravitational constant and $\theta$ is the angle between the BLR disk and the scattering region is assumed to be  $\theta\sim0$ in the case of equatorial scattering  and therefore, the BH  mass estimates is independent from the inclination. The constant $b$ is close to  0.5 for the dominant Keplerian-like motion.
One of the advantages is that this method can be applied to Mg\,II, C\,III], C\,IV broad lines, which would correspond to distant objects at high redshift, if these lines are observed in optical spectral range. In this work, we report the first spectropolarimetric observations of the Mg\,II line for a distant AGN SBS\,1419+538 and we compare the SMBH mass estimated using the \citetalias{2015ApJ...800L..35A} method with the estimates provided by different authors using other methods (most notably reverberation mapping).

\section{Polarimetric observations of M\lowercase{g} II spectral line}

In order to test the model of polarization changes for the Mg\,II line, in February 2019 we carried out the spetropolarimetric observations of the quasar SBS\,1419+538. SBS\,1419+538 (RA 14 21 06.9 Dec +53 37 45.2, J2000) is a bright quasar (16.8 mag in the g-sdss band) at the redshift $z = 1.862$ determined for the first time via the Second Byurakan Survey \citep{sbs2}. The SDSS spectra of the quasar show broad (FWHM $\sim$ \SI{5000}{\kilo\meter\per\second}) components of the MgII and CIII] in the optical range \citep{sbs3,2011ApJS..194...45S}.

SBS\,1419+538 was observed with the 6-m telescope BTA of SAO RAS with the focal reducer SCORPIO-2 \citep{2011BaltA..20..363A}. We used a \ang{;;1} slit and a volume phase holographic grating covering the \SIrange{5800}{9500}{\angstrom} range with a maximum at \SI{7350}{\angstrom} to obtain the spectrum images. Double Wollaston prism divided the image of the entrance pupil according to four polarization directions -- \ang{0} and \ang{90}, \ang{45} and \ang{135}. Then the parameters of the linear polarization and intensity - the Stokes parameters $Q$, $U$ and $I$  were obtained simultaneously and are equal:
\begin{alignat}{2}
  Q(\lambda) &= \frac{I_0(\lambda) - I_{90}(\lambda)K_Q(\lambda)}{I_0(\lambda) + I_{90}(\lambda)K_Q(\lambda)}, \\
  U(\lambda) &= \frac{I_{45}(\lambda) - I_{135}(\lambda)K_U(\lambda)}{I_{45}(\lambda) + I_{135}(\lambda)K_U(\lambda)}, \\
  I(\lambda) &= I_{0}(\lambda) + I_{90}(\lambda)K_Q(\lambda) + I_{45}(\lambda) + I_{135}(\lambda)K_U(\lambda),
\end{alignat}
\noindent where $K_Q$ and $K_U$ are the coefficients of the channel transmission, $I_{0}, I_{90}, I_{45}, I_{135}$ correspond to the different polarization directions. Using $K_Q$ and $K_U$ coefficients one can minimize the influence of variable atmospheric depolarization \citep[see][for more details]{2012AstBu..67..438A}. Then the polarization degree $P$ and polarization angle $\varphi$ are obtained from the following relations:
\begin{alignat}{2}
	  P(\lambda) &= \sqrt{Q^2(\lambda) + U^2(\lambda)}, \\
    \varphi(\lambda) &=\frac{1}{2}\arctan[U(\lambda)/Q(\lambda)] + \varphi_0,
\end{alignat}
\noindent where $\varphi_0$ is the zero point of polarization angle. To correct the device spectral sensitivity and to find $\varphi_0$ the non-polarized spectrophotometric and polarized standards were observed before the object. The polarimetric accuracy was up to variations of the atmospheric depolarization. Due to the high galactic latitude of the quasar ($b \sim \ang{60}$) the ISM polarization is neglected. The observations of the object were performed in a series of 16 frames with \SI{300}{\second} exposure times in order to make robust statistical estimations. The observational techniques and analysis method have been described in more details in several papers \citep[see e.g.][]{2012AstBu..67..438A,2014MNRAS.440..519A,2015ApJ...800L..35A,2019MNRAS.482.4985A} and will not be repeated.

\section{Results}
\begin{figure}[ht]
  \centering
  \includegraphics[width=0.95\columnwidth]{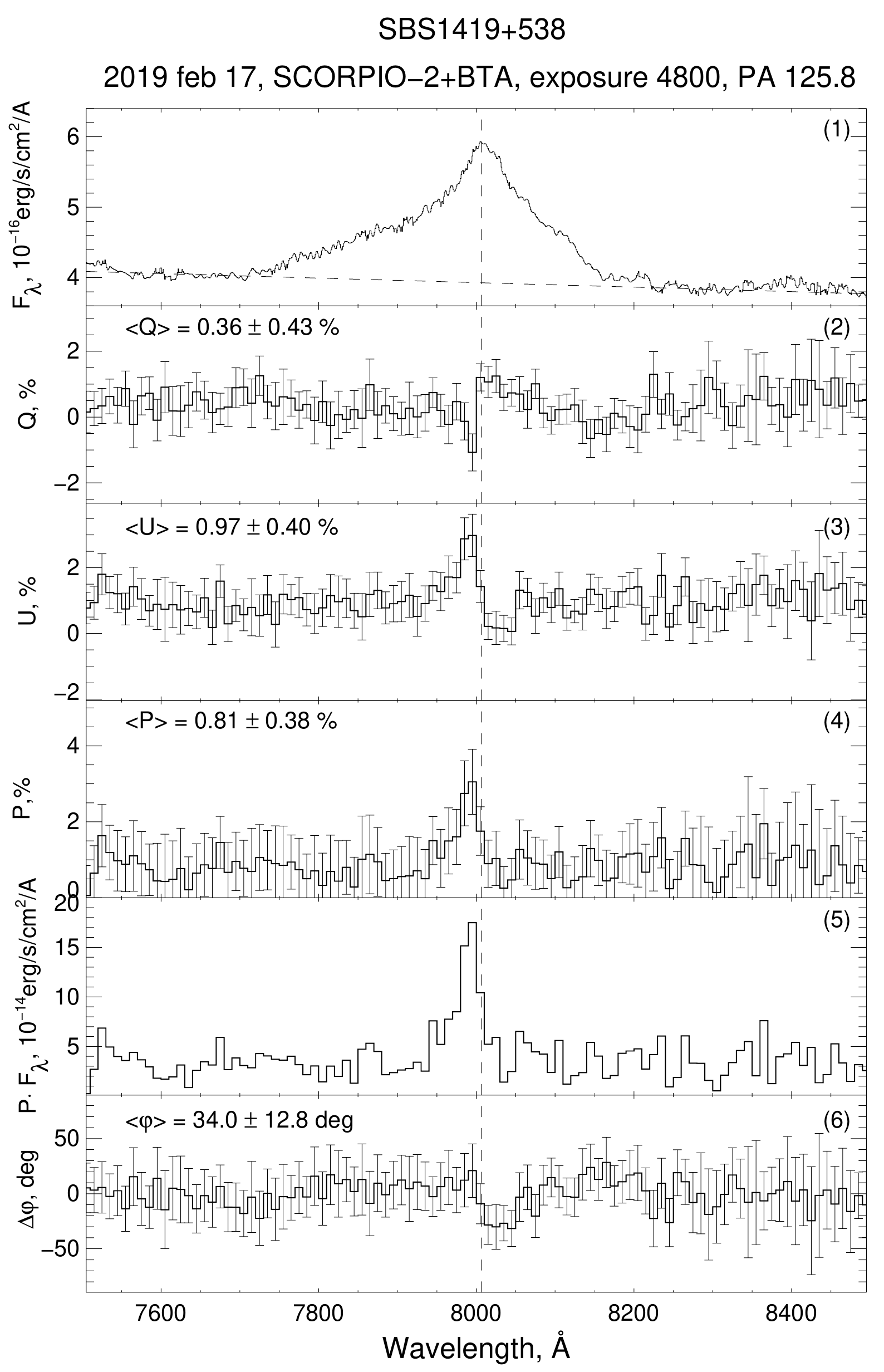}
  \caption{The Mg\,II spectral region (1st panel), the $Q$ and $U$ Stokes parameters (2nd and 3rd panels) and polarization degree and polarization angle (4th and 5th panel) across the line profile. }
   \label{f:obs1}%
\end{figure}

We extracted spectra and observed polarization parameters are shown in Fig.\,\ref{f:obs1}. The 1st panel shows the total flux in the spectral region near the broad Mg\,II line \SIrange{7500}{8500}{\angstrom} with 2{\AA} spectral resolution. The continuum emission is approximated here with a linear regression plotted with a dashed line. The 2nd and 3rd panels show the Stokes parameters $Q$ and $U$, respectively. The polarization degree $P$ and the polarization angle $\varphi$ are given on panels 4th and 5th. The Stokes parameters $Q$ and $U$, $P$ and $\varphi$ are binned over 10{\AA} and depend on the wavelength. For each bin, the value was calculated as a robust average in the 2-dimensional array of a size 10{\AA} by 16 exposures; the error bars are equal to the $1\sigma$ level as a robust standard deviation. A $2\sigma$ rejection threshold was used in order to avoid the influence of the outlier points (less than 1\% mostly due to the cosmic rays hints). The average values of the parameters $\langle Q \rangle, \langle U \rangle, \langle P \rangle, \langle \varphi \rangle$ are also given in the figure.

As the measured value of polarization is small and is comparable with the errors, the value of polarization degree $P$ is biased. The correction of $P$ to the bias was made according to the formula given in \citep{SS85}:
\begin{equation}
P = \sqrt{P^2_{\rm obs} - 1.41\sigma^2_P},
\end{equation}
where $P_{\rm obs}$ is measured value of polarization and $\sigma_P$ is its error. Therefore, there are unbiased values of $P$ given in Fig.\,\ref{f:obs1}.

The polarization profile of Mg\,II is single peaked and blue shifted may indicate some complex structure in the Mg II BLR, as e.g. outflowing/inflowing BLR \citep{2019MNRAS.484.3180P,2020MNRAS.497.3047S} or more complex as two component model \citep{2004A&A...423..909P} which can hide the expected two-peaks of the polarized profile in the case of disk-like motion \citep{2020MNRAS.497.3047S}. However in the case of pure disk-like motion, the single peaked polarized profile can be detected in the case of lower viewing inclinations \citep[see Fig.\,2][]{2020MNRAS.497.3047S}. Moreover, single peaked polarized line profile does not exclude dominant disk-like motion, and most of equatorial scattered type 1 AGNs in the sample of \citet{2019MNRAS.482.4985A} have single peaked polarized profile, but polarization angle swing indicates Keplerian like motion in the BLR \citep[see Figs.\,4-9][]{2019MNRAS.482.4985A}.

\begin{figure}[ht]
  \centering
  \includegraphics[width=0.99\columnwidth]{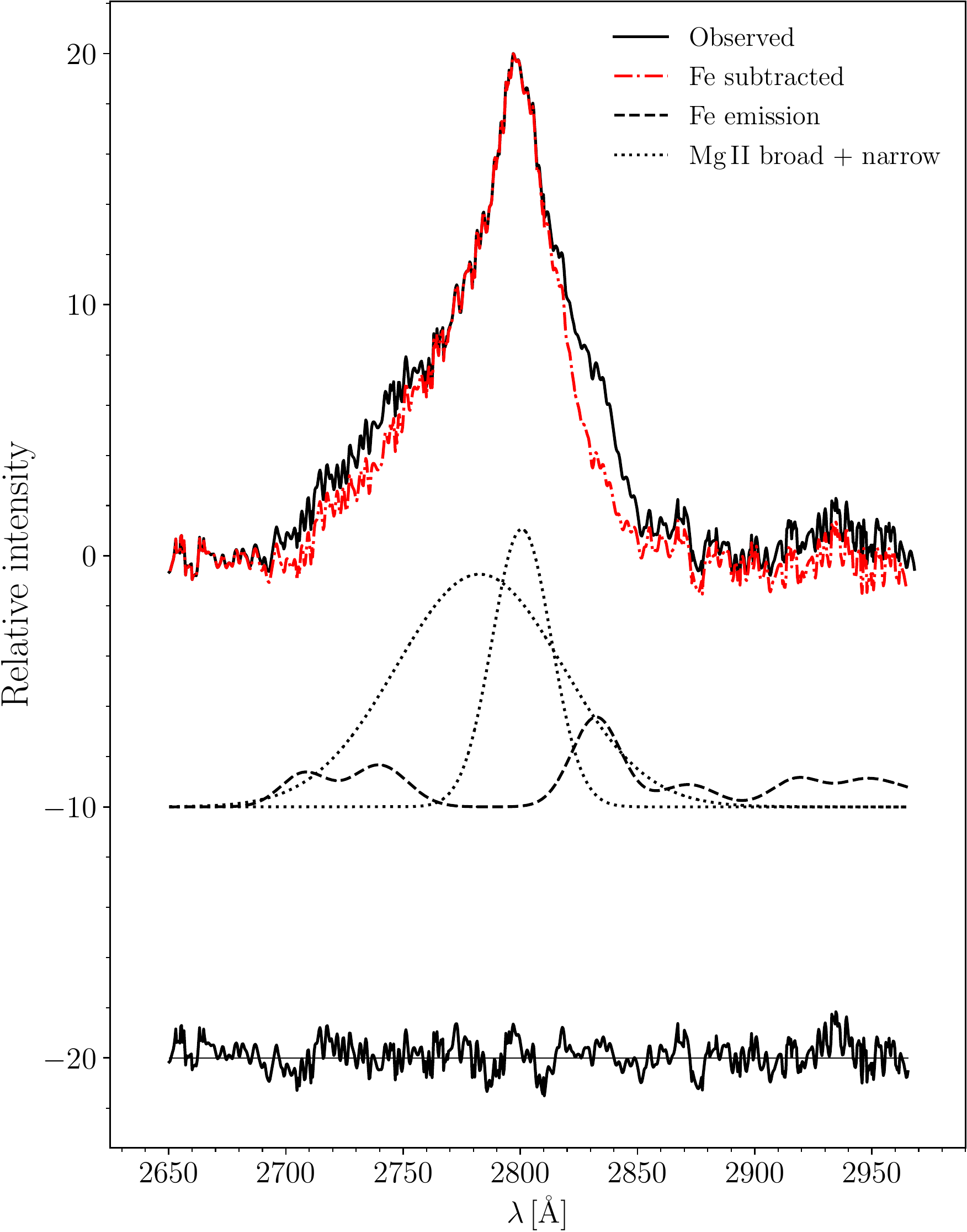}
  \caption{Decomposition of Mg\,II line emission. Solid black line denotes the observed spectrum; dashed red line is the Mg II profile after Fe II subtraction  (upper part). Middle part: Broad and narrow Gaussian components of the Mg\,II line (dotted line) with the contribution of the total UV Fe II emission (dashed black line); Residuals are shown at the bottom.}
   \label{f:Fe}%
\end{figure}

It is well-known that there is a strong iron emission underlying the Mg\,II line that also arises from the BLR. Estimation iron emission is a non-trivial task and much effort has been invested for solving this problem \citep[][and references therein]{2019MNRAS.484.3180P}. We used an improved model by \citet{2015ApJS..221...35K} that covers the spectral range between $\SI{2650}{}-\SI{3050}{\angstrom}$. Details regarding this model were extensively described by \citet{2019MNRAS.484.3180P}. An illustration of the Mg\,II decomposition is shown in Fig.\,\ref{f:Fe}. A blue asymmetry is dominant after Fe\,II subtracting indicating outflow, which is also seen in the blueshifted polarized profile.

To obtain the SMBH mass according to the polarization properties of the equatorially scattered emission in Mg\,II line by the method given in \citep{2015ApJ...800L..35A} one should estimate the radius of the scattering region $R_{\rm sc}$.  In \citet{2019MNRAS.482.4985A} the dependency connecting $R_{\rm sc}$ in AGN and the luminosity at \SI{1516}{\angstrom} was revealed:
\begin{equation}\label{af19}
\log\ind{R}{sc} = -(\SI{15.60 \pm 0.54}) + (\SI{0.40 \pm 0.01}) \log(\lambda L_{1516\AA}).
\end{equation}
As far as the spectropolarimetric observations given here have relatively bad photometric bounding due to the slit loses more confident estimations of luminosity of SBS\,1419+538 should be used. According to \citet{2016ApJ...818...30S} $\lambda L_{1350}=8.9 \cdot 10^{46}$ erg s$^{-1}$ and $\lambda L_{1700}=7.2 \cdot 10^{46}$ erg s$^{-1}$. As the continuum spectra slope in the spectral range is not steep let us consider $\lambda L_{1516} \approx 8 \cdot 10^{46}$ erg s$^{-1}$ and according to the dependency \ref{af19}, $R_{sc}$ is equal to:
\begin{equation} \label{sp}
\ind{R}{sc} = \SI{2041 \pm 683}{light days}.
\end{equation}
The error of \ind{R}{sc} was estimated by the bootstrapping method \citep{boot} and includes the errors of the coefficients from the Eqn.\,\ref{af19}. The asynchronism of the continuum luminosity taken from \citet{2016ApJ...818...30S} with respect to the spectropolarimetric observation from the given work and the $\lambda L_{1516\AA}$ uncertainty being smaller than the coefficients error were not taken into account.

\begin{figure}[ht]
  \centering
  \includegraphics[width=0.95\columnwidth]{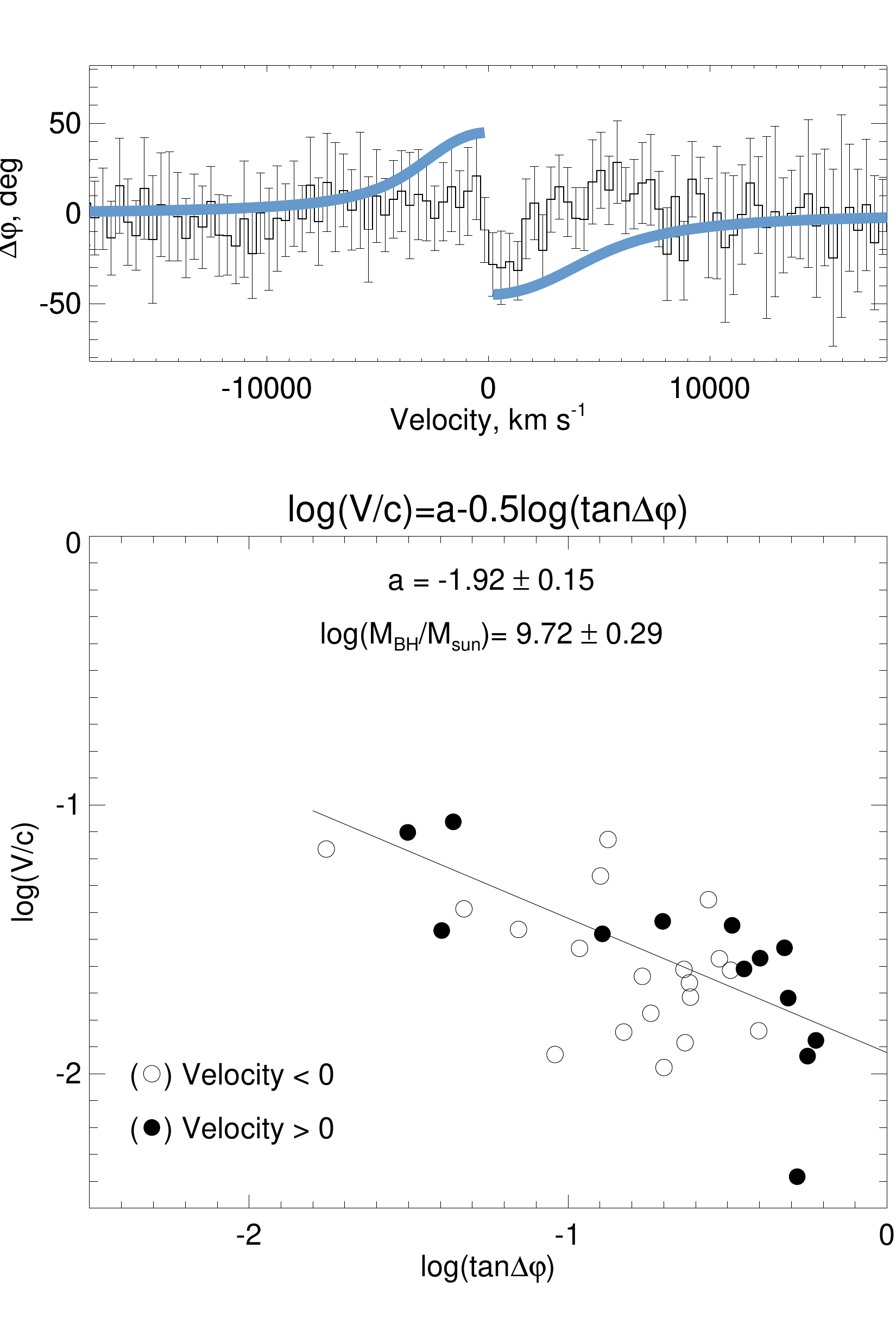}
  \caption{The relation $\log ( {V}/{c} )$ vs. $\log(\tan[\Delta \varphi])$ after rest frame correction. The velocity range for fitting is the interval between \SI{-d4}{\kilo\meter\per\second} and \SI{d4}{\kilo\meter\per\second}}.
   \label{f:obs2}%
\end{figure}

\noindent We applied the \citetalias{2015ApJ...800L..35A} method to find the black hole mass, and as it can be seen from Fig.\,\ref{f:obs2}, for Mg\,II line in the spectrum of SBS\,1419+538 the observational data could be fitted with a linear function with the regression coefficient $a = \SI{-1.95\pm0.13}{}$. Note here that the computed slope of $b$ coefficient is 0.46$\pm$0.11, so practically it was assumed identically equal to 0.5, which corresponds to the case of a Keplerian motion. Assuming that the BLR is co-planar with dust scattering region ($\cos^2(\theta) = 1$, see \citetalias{2015ApJ...800L..35A} for more details), we obtained that the SMBH mass is:

\begin{equation}
\log(\ind{\mathcal{M}}{bh}/\SI{}{\solarmass}) = \SI{9.67 \pm 0.27}{}.
\end{equation}

\noindent To examine the result we also tried other methods of indirect mass measurements. The mass could be estimated from the virial theorem. To calculate the virial product one should estimate the velocity dispersion of the broad emission line. We measured the FWHM = $\SI[multi-part-units=single]{4791\pm 552}{\kilo\meter\per\second}$ and line dispersion $\sigma = \SI[multi-part-units=single]{2275\pm 263}{\kilo\meter\per\second}$ after subtracting the Fe\,II contribution to the Mg\,II line using  the UV Fe II model given in  \citet{2019MNRAS.484.3180P}\footnote{models of the UV Fe\,II can be found at \url{http://servo.aob.rs/FeII_AGN/link7.html}}


The size of the BLR region in the Mg\,II line is estimated using the empirical BLR radius -- luminosity (R-L) relation \citep[see][]{2019ApJ...880...46C, 2020OAst...29....1P}. We used an updated R-L relation at \SI{3000}{\angstrom} given by \citet{2020ApJ...896..146Z}. The estimation of the quasar luminosity was obtained from \citep{2016ApJ...818...30S} $\lambda L_{3000}=\SI{6.1d46}{\erg\per\second}$. The BLR size was estimated as $R_{\rm BLR} = 1195_{-541}^{+936}$ light days. Therefore, the relation between the scattering region size and the BLR size is $R_{\rm sc}/ {R_{\rm BLR}} \approx 1.7 \pm 0.7$. This value is in a good agreement with the mean ratio obtained by \citet{2019MNRAS.482.4985A} as well as models by \citep{2018A&A...614A.120S} for which the ratio is expected to be in range between $1.5-2.5$.


The virial product could be calculated:
\begin{equation} \label{vp}
VP = \frac {\sigma^2 \ind{R}{BLR}} {G} \approx \SI{1.2d9}{\solarmass}.
\end{equation}




From the profile of the polarization angle, it is possible to determine the BLR direction of rotation. A maximum of the polarization angle in the blue wing of the line followed by the minimum in the red wing corresponds to the anticlockwise rotation of the central engine \citep{2018A&A...614A.120S}.

\begin{table*}[tbp]
\begin{center}
\caption{Previous measurements of SBS\,1419+538 found in literature. Columns from left to right: references, redshift, continuum luminosities at 1350; 1700; 3000, bolometric luminosity, full width at half maximum (FWHM) of the Mg\,II and C\,IV, the broad line region radius and estimated SMBH mass. We ignore the error bars that were not given by previous authors. All SMBH measurement were obtained by a single-epoch approach.}
\hspace*{-2cm}\resizebox{1.08\textwidth}{!}{\begin{tabular}{lcccccccccc}
\toprule\toprule
\multirow{2}{*}{Reference} & $z$ & $\log\lambda L_{1350}$ & $\log\lambda L_{1700}$ & $\log\lambda L_{3000}$ & $\log L_{\mathrm{Mg\,II}}$ & FWHM Mg\,II & FWHM C\,IV & $R_\mathrm{BLR}$ & $\log(\ind{\mathcal{M}}{bh})$ \\
&  & \SI{}{\erg\per\second} & \SI{}{\erg\per\second} & \SI{}{\erg\per\second} & \SI{}{\erg\per\second} & \SI{}{\kilo\meter\per\second} & \SI{}{\kilo\meter\per\second} & \SI{}{ld} & \SI{}{\solarmass}\\
\midrule
\citet{2008ApJ...680..169S} & 1.8583 & & & 46.89 & & 8557 & & & 10.168  \\     
\citet{2011ApJS..194...45S} & 1.8577 & $47.035\pm0.004$  & & $46.902\pm0.011$ & $44.97\pm0.01$ & $5889.7\pm821.6$ & $3352.5\pm106.5$ & & $10.08\pm0.15$ \\
\citet{2011ApJS..194...42R} &       1.8583 & & & 46.88 & & & & $813$ & 9.91 \\
\citet{2016ApJ...818...30S} & 1.863  & $46.9482\pm0.0022$ & $46.8588\pm0.0012$ & $46.7887\pm0.0004$ & \\
\citet{2019ApJ...887...38G} & 1.862  & $46.948\pm0.003$ & & & & & & & 9.31\\
This work                   & 1.862  & & & & $45.50\pm0.02$ & $4791\pm552$ & & $1195_{-541}^{+936}$ & $9.67\pm0.27$ \\

\bottomrule
\end{tabular}}
\label{t:objects}
\end{center}
\end{table*}

\section{Discussion and conclusions}
Magnesium lines are often associated with powerful outflows in addition to Keplerian motion \citep{2020NatAs.tmp..237L}. The outflows may be triggered by radiation pressure from the accretion disk and recently have been directly observed \citep{2020ApJ...904..149M}. In our previous works \citep{2018A&A...614A.120S,2020MNRAS.497.3047S}, we found that the \citetalias{2015ApJ...800L..35A} method may be used with sufficient accuracy even if the outflows are present. The main uncertainty in the SMBH mass estimate is proportional to the the radius of the scattering region for which we lack direct measurements. Instead, we rely on various scaling relations that typically involve measured UV, optical or infrared luminosity at certain wavebands \citep{2014ApJ...788..159K,2019MNRAS.482.4985A}, which in principle increase the error of the estimated SMBH mass. Another difficulty that arises using the current observational technique is the upper magnitude limitation. Due to the high redshift of the observed object, we are prone to observe only the brightest quasars in the spectropolarimetry mode. 


As follows from the description of the spectropolarimetric \citetalias{2015ApJ...800L..35A} method, the two main advantages of the approach are the use of single-epoch observations and independence from the orientation of the AGN relative to the observer. Due to the accumulated data on AGN reverberation mapping and the relatively high statistical accuracy of the luminosity dependences on the BLR size, the mass estimate can also be obtained from single observations. In Table \ref{t:objects} we report previous measurements of SBS\,1419+538 found in literature. Earlier SMBH estimates using data from SDSS campaign are close to $\log(\ind{\mathcal{M}}{bh}/\SI{}{\solarmass}) \approx \SI{10}{}$ \citep{2008ApJ...680..169S,2011ApJS..194...45S,2011ApJS..194...42R,2019ApJ...887...38G}. Generally, our results are in good agreement with the SMBH estimates. However, this estimate will depend on an unknown dimensionless factor $f \approx 1 \sim 10$, depending on the system orientation and geometry \citep{2000ApJ...533..631K,2004ApJ...615..645O,2013ApJ...767..149B}. Thus, the mass estimation error can reach 1 order of magnitude. A joint approach combining several types of mass estimation allows us to give more accurate and independent estimates of the masses.

A comparison of two independent estimates of the masses of the SMBH in Eqn.\,\ref{sp} and \ref{vp} allows us to estimate the dimensionless factor $f$, which in this case is equal to approximately 4. This value of the factor is 
close to the average value \citep[$f = 5.5$ is usually assumed for the most AGNs, see][]{2004ApJ...615..645O}.
Even if we assume that the $\ind{\mathcal{M}}{bh}$ obtained by spectropolarimetry is overestimated by 35\%, how is this it is assumed that according to the results of numerical modelling \citep{2020MNRAS.497.3047S}, the factor $f$ 
is expected to be equal to approximately 3. 
Note here that the 2 times difference between the mass estimate given in this article and in \citet{2019ApJ...887...38G} can also be explained by an incorrect choice of $f$.

An additional difficulty of the \citetalias{2015ApJ...800L..35A} method is the need to to estimate the inner radius of dusty torus where equatorial scattering is probably starting \citep[see Fig. 1 in][]{2020ApJ...892..118S}. Since there are no estimates of the radius of the dust torus in the IR range for the quasar SBS\,1419+538 in the literature, and the results of the SDSS RM campaign have not been published yet. Therefore, we used the empirical relation \ref{af19}. Undoubtedly, this worsens the accuracy of the estimate of the size of the $R_{\rm sc}$. However, the $R_{\rm sc}/R_{\rm BLR}$ ratio is close to the theoretically predicted, which indicates that the error in determining $R_{\rm sc}$ is not large than 30\%, i.e. lies within the accuracy of the \citetalias{2015ApJ...800L..35A} method for the Mg\,II line \citep{2020MNRAS.497.3047S}. In the future, the data of the SDSS-RM project or the results of the application of a new method for estimating the scattering region by reverberation mapping in polarized light \citep{2020ApJ...892..118S} will help to improve the accuracy of the mass estimate.

We apply for the first time the \citetalias{2015ApJ...800L..35A} method for Mg\,II broad line. Future work using the existing facility will include additional spectropolarimic observations of distant quasars focusing on C\,III] and C\,IV emission lines. Current limitations will be largely surpassed with a next-generation instrument POLLUX \citep{2018SPIE10699E..06M} that will be aboard the mission LUVOIR \citep[Large UV/Optical/IR Surveyor,][]{2019arXiv191206219T}.


\section*{Acknowledgements}
We dedicate this work to Viktor Leonidovich Afanasiev$^\dagger$\footnote{Deceased on December 21th 2020.} who performed observations and data reduction, as well set up the polarimetric observations at SAO that provides this measurements. This work was supported by the F.R.S.–FNRS under grant PDR T.0116.21; the Ministry of Education and Science Republic of Serbia through the project \textnumero451-03-68/2020-14/200002. Đ.\,Savić thanks the RFBR for the realization of the three months short term scientific visit at SAO funded by the grant \textnumero19-32-50009. E. Shablovinskaya thanks the grant of Russian Science Foundation project number \textnumero20-12-00030 "Investigation of geometry and kinematics of ionized gas in active galactic nuclei by polarimetry methods", which supported the spectropolarimetric data analysis.




\bibliographystyle{aasjournal}

\end{document}